\documentclass[reprint,
groupedaddress,
amsmath,amssymb,
aps,
longbibliography,
prl,
hidelinks]{revtex4-2}

\usepackage{graphicx}
\usepackage{soul}
\usepackage{dcolumn}
\usepackage{bm}
\usepackage[hidelinks]{hyperref}
\usepackage{xr-hyper}
\usepackage{xcolor}
\usepackage[normalem]{ulem}
\usepackage{amsmath,mathtools}
\usepackage{amssymb}
\usepackage{physics}
\usepackage{fancyhdr}
\usepackage{parskip}
\usepackage{lmodern}
\usepackage{graphicx} 
\usepackage{environ}

\usepackage{mathtools}
\newcommand\underrel[3][]{\mathrel{\mathop{#3}\limits_{%
			\ifx c#1\relax\mathclap{#2}\else#2\fi}}}

\begin{document}

\preprint{APS/123-QED}

\title{Nested Stochastic Resetting: Nonequilibrium Steady-states and Exact Correlations}
	\author{Henry Alston}
	\thanks{These authors contributed equally.}
	\affiliation{Department of Mathematics, Imperial College London, South Kensington, London SW7 2AZ, United Kingdom}
		
	\author{Callum Britton}
	\thanks{These authors contributed equally.}
	\affiliation{Department of Mathematics, Imperial College London, South Kensington, London SW7 2AZ, United Kingdom}
	
	\author{Thibault Bertrand}%
	\email{t.bertrand@imperial.ac.uk}
	\affiliation{Department of Mathematics, Imperial College London, South Kensington, London SW7 2AZ, United Kingdom}

\date{\today}

\begin{abstract}
\noindent Stochastic resetting breaks detailed balance and drives the formation of nonequilibrium steady states . Here, we consider a chain of diffusive processes $x_i(t)$ that interact unilaterally: at random time intervals, the process $x_n$ stochastically resets to the instantaneous value of $x_{n-1}$. We derive analytically the steady-state statistics of these nested stochastic resetting processes including the stationary distribution for each process as well as its moments. We are also able to calculate exactly the steady-state two-point correlations $\langle x_n x_{n+j}\rangle$ between processes by mapping the problem to one of the ordering statistics of random counting processes. Understanding statistics and correlations in many-particle nonequilibrium systems remains a formidable challenge and our results provide an example of such tractable correlations. We expect this framework will both help build a model-independent framework for random processes with unilateral interactions and find immediate applications, e.g. in the modelling of lossy information propagation.
\end{abstract}

\maketitle

Stochastic resetting, the return of a system to a pre-determined state at random times, leads to striking departures from conventional stochastic dynamics. At the single-particle level, diffusive motion interposed by resetting events, where the position of the particle is moved instantaneously to one of a predetermined set of sites, is a central model in the study of minimal processes breaking detailed balance due to its simplicity and analytical tractability \cite{Evans2011a, Evans2020}. It has been shown to generically drive the formation of nonequilibrium steady states \cite{Evans2011b, Eule2016, Mendez2016, Dibello2023, Roberts2024} and finds broad relevance in biological processes across length scales as a mechanism for reducing mean-first passage times in target search problems  \cite{Condamin2007, Benichou2005, Majumdar2021, DeBruyne2022, Redner2001, Janson2012, Reuveni2016, Pal2017, Evans2013, Bray2013, Faisant2021, Besga2020,Benichou2009, Benichou2011, Bressloff2020, Alston2024}. 

Yet, the role of resetting in many-body systems with interacting particles remains largely unexplored. A small number of recent works have studied the impact that resetting mechanisms can have in a variety of many-particle systems \cite{Biroli2023, Nagar2023, Meigel2023, Bressloff2024}. In particular, resetting events can serve as a way to introduce and control correlations in systems without explicit interactions between processes \cite{Biroli2023}. These correlations are central to our analysis of collective behaviours in complex systems including structural changes and emergent phenomena. Particle systems for which correlations between particles can be calculated exactly remain elusive: these tractable examples provide both physical insight and potential avenues for constructing a more general framework. 

Building upon the analytic formalism developed for resetting processes, we introduce in this Letter a novel many-particle model that remains analytically solvable in steady-state despite the existence of strong correlations arising from its dynamics. We consider a chain of diffusive processes with unilateral interactions between nearest-neighbours on the chain enforced by a pairwise resetting mechanism. We demonstrate that this is sufficient to generate correlations between particles that are arbitrarily far apart in the chain, calculating exactly the steady-state correlations in particle positions of the form $\langle x_n x_{n+j}\rangle$ by reconstructing the problem as one on the ordering of random counting processes. Our work both develops a framework for evaluating correlations in systems with unilateral interactions, while also further illuminating the role resetting mechanisms can play in inducing correlations in many-particle systems. 

\textit{Nested stochastic resetting. ---} We consider a system of $N$ Brownian particles confined to the real line: $x_n(t)\in \mathbb{R}$ for $n\in\{1, \dots, N\}$. The diffusive motion of each particle is independent and characterised in each case by the diffusion coefficient $D$. Each particle is independently subjected to resetting events which occur with rate $r$ in a Poissonian manner. In the event where $x_n$ undergoes a reset at time $t$, the position of $x_n$ is set to $x_{n-1}(t)$, that is, the instantaneous position of the ($n-1$)-th particle in the chain. When particle $x_1$ resets, it does so to the fixed time-independent distribution of positions $P_0(x)$.

We define the probability distribution for the position of the $n$-th particle as $P_n(x, t)=\mathbb{P}(x_n(t)=x)$. For all $n>0$, the Fokker-Planck equation describing the evolution of $P_n(x, t)$ takes the form 
\begin{equation}\label{eq:fp}
\partial_t P_n(x, t) = D\partial_{x}^2 P_n(x, t) - r P_n(x, t) + r P_{n-1}(x, t).
\end{equation}
The right-hand side consists of both a diffusive term and gain-loss terms stemming from the resetting process. It is straightforward to confirm that the total probability is conserved. For the remainder of this work, we assume that $P_0(x)$ has zero mean $x_0 = 0$ and a variance $\sigma_0$. Our results can be mapped on to the case of finite $x_0$ by working with re-defined variables of the form $y_n(t) = x_0 + x_n(t)$. In particular, $\langle y_n y_{n+j}\rangle = x_0^2 +  \langle x_n x_{n+j}\rangle.$

\textit{Nonequilibrium steady-state distributions. ---} We denote the steady-state distribution of particle $x_n$ as $P_n^*(x)=\lim_{t\rightarrow\infty}P_n(x, t).$ From the Fokker-Planck equation above, we derive a hierarchy of equations for the distributions $P_n^*(x)$ by solving Eq.\,\eqref{eq:fp} at steady-state: 
\begin{equation}\label{eq:fp_ss_itr}
0 = D \partial_x^2 P_n^*(x) - rP_n^*(x)  + r P_{n-1}^*(x).
\end{equation}
We define the Fourier transform of a function $f(x)$, which we will denote by $\hat{f}(q)$, and its inverse transform, respectively, as 
\begin{equation}
\hat{f}(q) = \int _{-\infty}^\infty dx\, f(x)e^{-i q x},\quad f(x) = \frac{1}{2\pi} \int_{-\infty}^\infty \hat{f}(q) e^{iqx}.
\end{equation}

\begin{figure}
\centering
\includegraphics[scale=1]{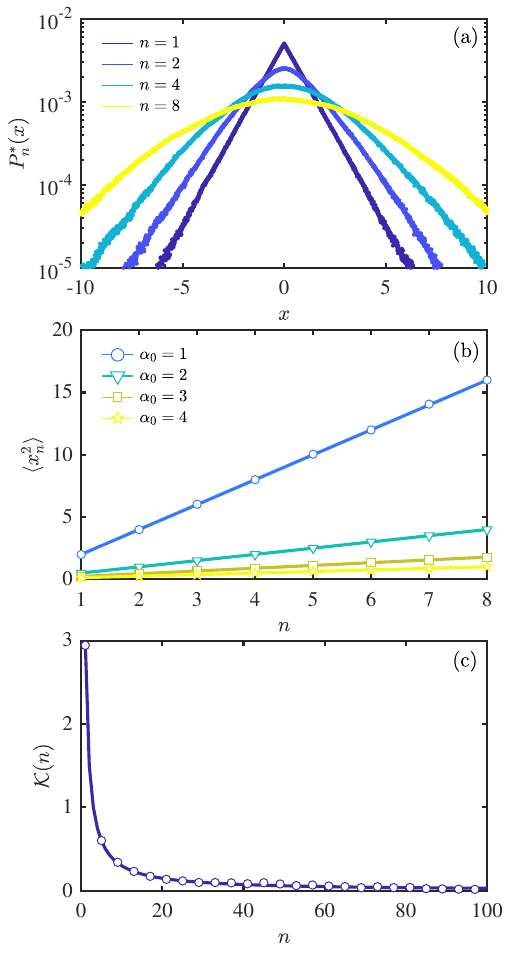}
\caption{\textit{Nonequilibrium steady-states and positional moments ---}  We compare our analytic results (solid line) for (a) the stationary distribution of particle $x_n$, $P_n^*(x)$ [see Eq.\,\eqref{eq:pn*_fs}], and (b) the positional variance of particle $x_n$ position, $\langle x_n^2\rangle$ [see Eq.\,\eqref{eq:variance}], to numerical simulations of the microscopic process (symbols), for $n=\{1, 2, 4, 8\}$ demonstrating the validity of our exact result (solid line). (c) Excess kurtosis $\mathcal{K}(n)$ converging to zero in the limit of large $n$ values (solid line - analytics and symbols - numerics).}
\label{fig:ness}
\end{figure}

Taking the Fourier transform of Eq.\,\eqref{eq:fp_ss_itr}, we derive an iterative formula for the distribution functions in Fourier space: we define $\alpha=\sqrt{r/D}$ and derive 
\begin{equation}\label{eq:pn_itr}
\hat{P}_n^*(q) = \frac{\alpha^2}{q^2 + \alpha^2} \hat{P}_{n-1}^*(q)
\end{equation} 
which by recurrence leads for all $n$ to
\begin{equation}\label{pn*_fs}
\hat{P}_n^*(q) = \hat{P}_0(q) \left(\frac{\alpha^2}{q^2 + \alpha^2}\right)^n.
\end{equation}
Finally, we derive our result for $P_n^*(x)$ by taking the inverse Fourier transform of $\hat{P}_n^*(q)$ (see \cite{Bateman1954} for instance) leading to the following convolution in real-space:
\begin{equation}\label{eq:pn*_fs}
P_n^*(x) = \frac{\alpha}{\sqrt{\pi}(n-1)!}\biggl(\frac{\alpha|x|}{2}\biggr)^{n-\frac{1}{2}}K_{n-\frac{1}{2}}(\alpha |x|) * P_0(x)
\end{equation}
where $K_m$ is a modified Bessel function of the second kind. From this, we conclude that for all finite values of $n$, no matter how large, particles have a well-defined nonequilibrium steady-state distribution with finite variance [see Fig.\,\ref{fig:ness}(a)]. We remark that $P_1^*(x)$ is exactly 
\begin{equation}
P_1^*(x) = \frac{\alpha}{2} \exp(-\alpha |x|) * P_0(x)
\end{equation}
which is the stationary distribution known for a particle undergoing Brownian motion with Poissonian resetting to a fixed distribution $P_0(x)$ \cite{Evans2011a, Evans2020}. In what follows, we take $P_0(x) = \delta(x)$ for illustrative purposes but our results are easily generalizable. 

\textit{Positional moments. ---} We now calculate the moments of each particle's position. First, we note that the odd moments vanish by symmetry as $x_0=0$ and the dynamics are invariant under reflection ($x\rightarrow -x$). The remaining moments can be calculated directly from the expressions derived above for the stationary probability distributions.

Calling again on Ref.\,\cite{Bateman1954} (see \footnote{See Supplemental Material at [] for further analytical and computational details, which includes Refs. []} for details), we can derive the following closed-form expressions for the moment of order $2q$ of the $n$-th particle in the nested stochastic resetting process:
\begin{equation}
\langle x_n^{2q}\rangle = \int_{-\infty}^\infty dx \: x^{2q} P_n^*(x) = {n+q-1\choose n-1}\frac{(2q)!}{\alpha^{2q}}. 
\label{eq:moments}
\end{equation}

In particular, the variance of particle $x_n$ takes the form 
\begin{equation}
\langle x_n^2 \rangle = \frac{2n}{\alpha^2} = n\langle x_1^2 \rangle,
\label{eq:variance}
\end{equation}
a surprisingly simple result: the mean squared displacement from the origin grows linearly with the distance along the resetting chain [as can be seen in Fig.\,\ref{fig:ness}(b)]. 

As seen in Fig.\,\ref{fig:ness}(a), the steady-state distribution crosses over from double exponential tails to Gaussian tails as $n$ increases. To confirm this, we measure the excess Kurtosis, defined as
\begin{equation}
\mathcal{K}(n) = \frac{\langle x_n^4\rangle}{\langle x_n^2\rangle ^2} - 3.
\end{equation}
and using Eq.\,(\ref{eq:moments}), we obtain 
\begin{equation}\label{eq:kurtosis}
\mathcal{K}(n) = \frac{3}{n},
\end{equation}
which shows that the excess Kurtosis indeed converges to $0$ when $n \to \infty$ [see Fig.\,\ref{fig:ness}(c)]. Note that $\mathcal{K}$ is a measure of non-Gaussianity; indeed, we have $\mathcal{K}\equiv 0$ for a Gaussian distribution. Our results thus imply that the stationary distributions for $x_n$ converge towards a Gaussian distribution with zero-mean and variance given by Eq.\,(\ref{eq:variance}) as $n$ increases.

\begin{figure}[t!]
    \centering
    \includegraphics[scale=1]{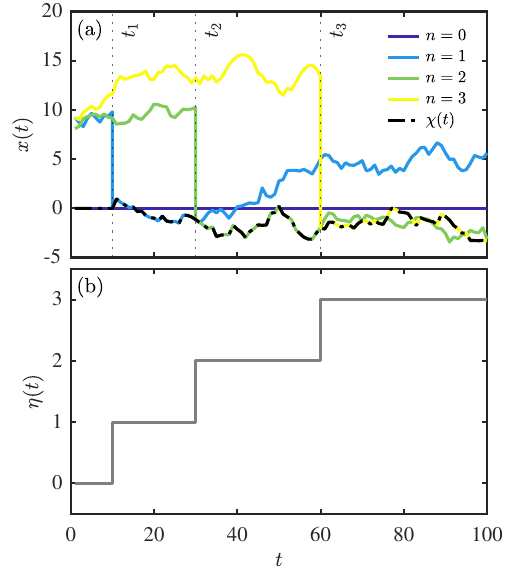}
    \caption{\textit{Mapping to diffusive trajectory} --- (a) Example realization of the nested resetting process for $N=3$. The diffusive trajectory $\chi(t)$ constructed from the coupled microscopic dynamics as instructed from Eq.\,\eqref{eq:Xdef} is shown as a black dashed line. (c) The associated index from this trajectory, $\eta(t)$, such that $\chi(t) = x_{\eta(t)}(t)$ for all $t$.}
    \label{fig:mapping}
\end{figure}

\textit{Static two-point correlation functions. ---} Going further, we now compute exactly the two-point correlations of the form $\langle x_{n} x_{n+j} \rangle$, where $n$ and $j$ are positive integers. To find an expression for this correlator, we recognise that $x_n(T)$ at some large time $T$ can be understood as the displacement of a purely diffusive particle over $n$ exponentially distributed waiting times. Indeed, at large time $T$ (such that the initial conditions have been forgotten and the system is in its nonequilibrium steady-state), we know that $x_n$ last reset to the position of $x_{n-1}$ at some time $t_n=T-\tau_n$, where $\tau_n$ is an exponentially distributed random variable with rate $r$ as dictated by the statistics of the resetting mechanism. In the time interval $t \in (t_n, T]$, the motion of $x_n(t)$ is entirely diffusive. At time $t_n$, we have that $x_n(t_n)=x_{n-1}(t_n)$. Similarly, we can identify a time $t_{n-1}=T-\tau_n-\tau_{n-1}$ (where $\tau_{n-1}$ is independent of $\tau_n$ but identically distributed) such that the motion of $x_{n-1}$ is entirely diffusive over $t_{n-1} < t \le t_n$. This process can be iterated until reaching $x_0=0$, which happens at time $t_1=T-\sum_{m=1}^n \tau_m$ (see Fig\,\ref{fig:mapping}). 

Following this procedure, we can construct (for any realization of the resetting events) an entirely diffusive trajectory that originates at $0$ and arrives at $x_n(T)$ at time $T$ which we denote by $\chi(t)$: 
\begin{equation}\label{eq:Xdef}
\chi(t) = 
\begin{cases}
    0, & t \leq t_1 \\
    x_{m}(t), & t \in (t_m, t_{m+1}],~(1\le m\le n-1)\\
    x_n(t), & t \in (t_n, T]
\end{cases}
\end{equation}
In doing so, we also define the counting process $\eta(\tau)$ which describes the indices associated with the progression of the diffusive trajectory through resetting events:
\begin{equation}\label{eq:Ndef}
\eta(t) = 
\begin{cases}
    0 & t \leq t_1 \\
    m & t \in (t_m, t_{m+1}],~(1\le m\le n-1)\\
    n & t \in (t_n, T].
\end{cases}
\end{equation}
Using these definitions, we write $\chi(t)=x_{\eta(t)}(t)$ or equivalently, $x_n(t)=\{\chi(t), \eta(t)\}$. In \cite{Note1}, we show that we can easily rederive results (\ref{eq:pn*_fs}) and (\ref{eq:moments}) in this framework.

Now consider the two processes $x_n(t)=\{\chi(t), \eta(t)\}$ and $x_{\bar{n}}(t)=\{\bar{\chi}(t), \bar{\eta}(t)\}$, with $\bar{n} = n+j$, $j>0$. By definition of $\chi$ and $\eta$, we conclude that if $\eta(t) = \bar{\eta}(t) \equiv m$ for some time $t<T$, then the two diffusive processes are at the same location at $t$, leading to $\chi(t) = \bar{\chi}(t)$. We also argue that in this case, the jumping events of the two processes $N$ and $\bar{N}$ are synchronised before $t$, thus the processes are identical for all $t'\leq t$. This implies that the static correlator (i.e. in the limit of large $T$) between two particles on the resetting chain is entirely controlled by $t_\ell$ the last time before $T$ such that the $\eta(t_\ell) = \bar{\eta}(t_\ell)$. Note that for $t\in(t_\ell, T]$, $\chi(t)$ and $\bar{\chi}(t)$ are uncorrelated as $\eta(t)\ne \bar{\eta}(t)$.

In \cite{Note1}, we show that using this mapping we can write the correlator in the form
\begin{equation}
    \langle x_n x_{n+j} \rangle = \sum_{m=1}^{n} \alpha_{m}^{n,n+j} \langle x_m^2 \rangle,
    \label{eq:corr_expr}
\end{equation}
where the coefficients are given by the following probabilities
\begin{equation}
    \alpha_{m}^{n, n+j} \equiv \mathbb{P}(n_\ell = m | \{n, n+j\} ).
\end{equation}
where $\mathbb{P}(n_\ell = m | \{n, n+j\} )$ is the probability that $n_\ell = m $ given that the two particles of interest are $n$ and $\bar{n}= n +j$. Said differently, $\alpha_m^{n, n+j}$ represents exactly the probability that $\eta(t) = \bar{\eta}(t) = m$ at some time $t$, but also crucially that there is no time $t'$ for which $\eta(t')=\bar{\eta}(t')=m'$ for some $m' \in \{m+1, \dots, n-1, n\}$.

\begin{figure}[t!]
    \centering
    \includegraphics[scale=1]{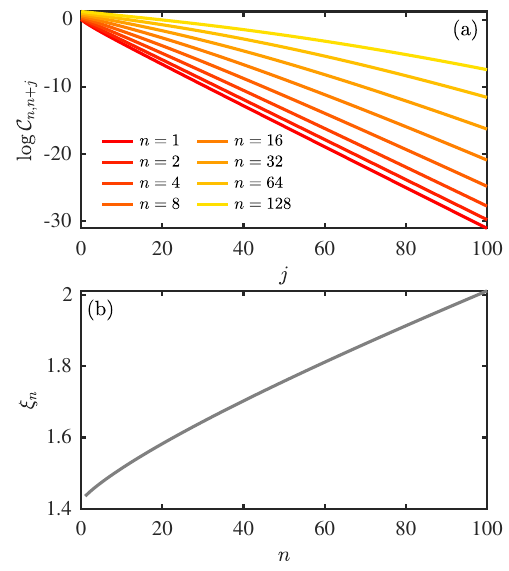}
    \caption{{\it Re-scaled correlation function $\mathcal{C}_{n, n+j}$} --- (a) analytical re-scaled correlation function $\mathcal{C}_{n, n+j}$, as defined in Eq.\,\eqref{eq:resclcorr} showing an exponential decay at large $j$ values of the form $\mathcal{C}_{n, n+j}\sim \exp (-j/\xi_n)$. (b) Correlation length $\xi_n$ extracted by fitting the curves in (a). }
    \label{fig:scaling}
\end{figure}

The problem of calculating $\alpha_{m}^{n, n+j}$ can be mapped to a discrete-time random walk (DTRW) on a 1D lattice, where two particles start at sites $n$ and $n+j$ and at each step, one of the two particles is chosen to make a step to the left with equal probability (until it reaches $0$ at which point it stops). Eventually, the particle initially at $n+j$ will hop on the site occupied by the particle initially at $n$: the probability that this happens at site $m$ for the first time is exactly $\alpha_{m}^{n, n+j}$ and is written 
\begin{subequations}
\begin{align}
\alpha_n^{n, n+j} &= \mathcal{B}(j, j) =  2^{-j}, \label{eq:alpha_a}\\
\alpha_{n-k}^{n, n+j} &= \mathcal{B}(k, j+2k) - \sum_{l=1}^{k} \alpha_{n-k+l}^{n, n+j} \mathcal{B}(l, 2l) .\label{eq:alpha_b}
\end{align}
\label{eq:alpha_coeff}%
\end{subequations}
where $\mathcal{B}(m, n) = 2^{-n} {n \choose m}$ as shown in \cite{Note1}. We can intuitively understand Eq.\,(\ref{eq:alpha_coeff}) as follows: Eq.\,(\ref{eq:alpha_a}) states that the two particles meet at $n$ only if the particle at $n+j$ hops exactly $j$ times; further, Eq.\,(\ref{eq:alpha_b}) states that to meet at $n-k$, the particles must meet at $n-k$ (first term) without having met first at another site $n-k'$ where $k'<k$ (second term).

Importantly, Eq.\,\eqref{eq:alpha_coeff} provides an exact, fully-analytic expression for all coefficients $\alpha_m^{n, n+j}$ and thus exact results for the correlations of the form $\langle x_n x_{n+j}\rangle$ through Eq.\,\eqref{eq:corr_expr} which constitutes one of our main results. We note that while in principle these correlations are tractable through the so-called last renewal approach (see details in \cite{Note1}) calculations quickly become arduous for increasing $n$ and $j$; the present approach circumvents this issue and is thus more generally applicable. 

As seen in Fig\,\ref{fig:scaling},  the re-scaled correlation function $\mathcal{C}_{n, n+j}$ defined as 
\begin{equation}\label{eq:resclcorr}
    \mathcal{C}_{n, n+j} = \frac{\langle x_{n}x_{n+j}\rangle}{\sqrt{\langle x_n^2\rangle \langle x_{n+j}^2\rangle}}
\end{equation}
decays exponentially for large enough value of $j$ taking the form $\mathcal{C}_{n, n+j}\sim \exp(-j/\xi_n)$, where $\xi_n$ represents the effective correlation length in terms of distance along the chain of resetting processes. Counterintuitively, we show that $\xi_n$ increases with $n$. In other words, particles in the chain of resetting processes separated by $j$ particles, say particles $x_n$ and $x_{n+j}$, become \textit{more} correlated as $n$ increases. 

To understand this result, considering the evolution of the stationary distributions for the re-scaled variables with zero-mean and unit-variance: $z_n = x_n/\sqrt{\langle x_n^2\rangle}$. The fact that $\xi_n$ increases with $n$ suggests that the stationary distributions for $z_n$ and $z_{n+j}$ themselves become more correlated as $n$ increases. To ascertain this convergence in the distributions of the re-scaled variables, we quantify the relative difference in excess Kurtosis between particles $n$ and $n+j$ in the chain. Using Eq.\,(\ref{eq:kurtosis}), we derive the difference in the excess Kurtosis, written as 
\begin{equation}
    |\mathcal{K}(n+j)-\mathcal{K}(n) |= \frac{3j}{n(n+j)}
\end{equation}
which decreases monotonically with $n$, as expected.

\textit{Discussion \& Outlook. ---} In this Letter, we define what we call the nested stochastic resetting process and study its steady-state properties obtaining a number of exact analytical results including probability distributions for the position of all particles in the resetting chain, moments of the distributions as well as static two-point correlators. Exact correlations for many-particle systems with broken detailed balance at the microscopic scale are often untractable and crucially missing from the literature. Altogether, we provide here an example of such tractable correlations for a many-body interacting resetting process for which correlations can be evaluated exactly. To do so, we exploited the unidirectional nature of the interactions to construct effective diffusive trajectories as dynamical descriptions of the position of the $n$-th particle in the chain.

The nested stochastic resetting process described here relies on a succession of particles probing the state of the previous particle down the chain at randomly distributed times. Stepping away from the classical picture of resetting as a spatial process, resetting events here can be seen as a noisy process reading out the state of their diffusive neighbor in real time leading to a copy of information at randomly selected times. At an intuitive level, the resetting events can thus be seen as information propagation events in which the original information represented by the distribution $P_0(x)$ gets transmitted in a lossy way down the chain in a process akin to that of word of mouth. When information spreads by word of mouth, each retelling introduces the potential for distortion or loss of detail (here represented by the particles' diffusion), especially as the message passes through more intermediaries. Our main results shows that paradoxically, those furthest down the chain—the least informed—can be the loudest (i.e. those with the furthest reach). In a society, where information is often propagated from individuals to individuals or communities to communities on social networks rather than via centralized news distribution channels, confidence in incomplete or misunderstood ideas can replace nuance and exactness in the information with oversimplification or sensationalism. In a sense, our results underscore the risks of amplification without verification in informal information networks.

Future work will seek to address the dynamical properties of nested resetting processes. Indeed, resetting mechanisms ensure a finite mean-first passage time for diffusive search processes \cite{Evans2011a, Benichou2005, Majumdar2021, DeBruyne2022, Redner2001, Janson2012, Reuveni2016, Pal2017, Evans2013, Bray2013, Faisant2021, Besga2020,Benichou2009, Benichou2011, Bressloff2020, Alston2024}. One natural question to ask would be how these first-passage times vary across particles in our system. Similarly, the extreme value statistics of coupled processes pose a formidable challenge for our framework of correlated random variables \cite{Hansen2020, Majumdar2020}. Finally, the breakdown of time reversal symmetry in microscopic (Markovian) processes is quantified through the evaluation of the entropy production rate \cite{ Kurchan1998, Gaspard2004, Peliti2021, Pruessner2022, Cocconi2020}. This diverges for the classical single-particle stochastic resetting process when there is only one resetting site. Recent work has considered extensions to these processes that ensure finite measurements of the entropy production (ensuring a physically relevant dynamical system) \cite{Mori2023, Alston2022b}. Analysis of the current model in this context, in particular for $n>1$, may offer one system in which calculating the exact many-body entropy production is not out of reach.

\begin{acknowledgments}
We thank Paul C. Bressloff for useful discussions. HA and CB were supported by a Roth PhD scholarship funded by the Department of Mathematics at Imperial College London. 
\end{acknowledgments}


%

\end{document}


\title{Supplemental Material for ``Nested Stochastic Resetting: Nonequilibrium Steady-states and Exact Correlations"}
	
	\author{Henry Alston}
	\thanks{These authors contributed equally.}
	\affiliation{Department of Mathematics, Imperial College London, South Kensington, London SW7 2AZ, United Kingdom}
		
	\author{Callum Britton}
	\thanks{These authors contributed equally.}
	\affiliation{Department of Mathematics, Imperial College London, South Kensington, London SW7 2AZ, United Kingdom}
	
	\author{Thibault Bertrand}%
	\email{t.bertrand@imperial.ac.uk}
	\affiliation{Department of Mathematics, Imperial College London, South Kensington, London SW7 2AZ, United Kingdom}
	

	\maketitle
 
 	\hrule
	\tableofcontents
 	\vspace{2em}
	\hrule

\section{Further details for the derivation of steady-state distributions and positional moments} 
\label{sec:bateman}

\subsection{Steady-state distributions $P^*_n(x)$}

In Eq.\,(5) of the main text, we identify the steady-state probability distribution in Fourier space. The steady-state distribution in real space is thus obtained as the following convolution
\begin{align}
 P_n^*(x) = \mathcal{I}_n(x) * P_0(x)
\end{align}
with $\mathcal{I}_n(x)$ defined as the following integral
\begin{equation}
\mathcal{I}_n(x) = \frac{1}{2\pi} \int_{-\infty}^\infty dq \: \left(\frac{\alpha_0^2}{q^2 + \alpha_0^2}\right)^n e^{iqx}.
\end{equation}

To evaluate this integral and find a result in real space, we use the following result (see for instance \cite{Bateman1954a} --- Sec.\,1.12, p.\,49)  
\begin{equation}
    \int_0^\infty dx\,x^{\pm \mu} K_{\mu} (ax) \cos(xy)  = \frac{\sqrt{\pi}}{2} (2a)^{\pm \mu} \Gamma(\pm\mu + 1/2)(y^2 + a^2)^{\mp \mu - 1/2},
\end{equation}
which can be manipulated into the following form
\begin{align}
    \int_0^\infty dx\,x^{\pm \mu} K_{\mu} (ax) \cos(xy)  =  \frac{1}{2} \int_{-\infty}^\infty dx |x|^{\pm \mu} K_{\mu} (a|x|) e^{-ixy},
\end{align}
allowing us to identify the following Fourier transform
\begin{equation}
    \int_{-\infty}^\infty dx |x|^{\pm \mu} K_{\mu} (a|x|) e^{-ixy}  = \sqrt{\pi} (2a)^{\pm \mu} \Gamma(\pm\mu + 1/2)(y^2 + a^2)^{\mp \mu - 1/2},
\end{equation}
We therefore use the Fourier inversion theorem, with $\mu = n-1/2, a = \alpha$, and arrive at
\begin{equation}
    P_n^*(x) = \frac{\alpha}{\sqrt{\pi}(n-1)!}\biggl(\frac{\alpha|x|}{2}\biggr)^{n-1/2}K_{n-1/2}(\alpha |x|) * P_0(x).
\end{equation}

\subsection{Positional moments $\langle x^{2q}_n\rangle$}

The moment of order $2q$ of the $n$-th particle in the nested stochastic resetting process is given by 
\begin{align}
    \langle x_{n}^{2q} \rangle = \int_{-\infty}^{\infty} dx\, x^{2q} P_{n}^*(x).
\end{align}
From \cite{Bateman1954b} (Sec.\,10.2, p.\,127), we write the following relation
\begin{equation}
    \sqrt{a}\int_0^\infty dx\,x^{\rho-1/2}K_\nu(a x) = 2^{\rho-3/2}a^{-\rho} \Gamma\left(\frac{\rho}{2} + \frac{\nu}{2} + \frac{1}{4}\right) \Gamma\left(\frac{\rho}{2} - \frac{\nu}{2} + \frac{1}{4}\right),
\end{equation}
and upon identifying $\rho = 2q + n, a=\alpha, \nu = n-1/2$ we obtain
\begin{align}
    \sqrt{\alpha}\int_0^\infty dx\,x^{2q+n-1/2}K_{n-1/2}(\alpha x) &= 2^{2q+n-3/2}a^{-(2q+n)} \Gamma\left(q+n\right) \Gamma\left(q + \frac{1}{2}\right).
\end{align}
Rewriting the integral as 
\begin{align}
   \int_0^\infty dx\,x^{2q+n-1/2}K_{n-1/2}(\alpha x) &= \frac{1}{2}\int_{-\infty}^\infty dx\,x^{2q} |x|^{n-1/2}K_{n-1/2}(\alpha |x|),
\end{align}
we arrive at the final result
\begin{equation}
    \langle x_n^{2q}\rangle =\frac{1}{(n-1)!}\frac{1}{\sqrt{\pi}}\biggl(\frac{2}{\alpha}\biggr)^{2q}\Gamma(q+1/2)\Gamma(n+q) = \frac{(n+q-1)!(2q)!}{(n-1)!q!\alpha^{2q}} = {n+q-1\choose n-1}\frac{(2q)!}{\alpha^{2q}}. 
\end{equation}

\section{Deriving steady-state distribution using the mapping to diffusive trajectories}
\label{sec:difftraj}

In the main text, we have shown via direct integration that the steady-state distributions were given by
\begin{equation}
P_n^*(x) = \frac{\alpha}{\sqrt{\pi}(n-1)!}\biggl(\frac{\alpha|x|}{2}\biggr)^{n-\frac{1}{2}}K_{n-\frac{1}{2}}(\alpha |x|) * P_0(x)
\end{equation}
where $K_m$ is a modified Bessel function of the second kind. In this section, we show that we can also derive this result for the stationary distribution $P_n^*(x)$ by using the mapping to diffusive trajectories introduced in the main text. 

For completeness, recall that we recognize that $x_n(T)$ at some large time $T$ can be understood as the displacement of a purely diffusive particle over $n$ exponentially distributed waiting times.
From this, we can construct (for any realization of the resetting events) a diffusive trajectory $\chi(t)$ that originates at $0$ and arrives at $x_n(T)$, which is defined as
\begin{equation}
\chi(t) = 
\begin{cases}
    0, & t \leq t_1 \\
    x_{m}(t), & t \in (t_m, t_{m+1}],~(1\le m\le n-1)\\
    x_n(t), & t \in (t_n, T]
\end{cases}
\label{eq:chi_def}
\end{equation}
Recall that this led to defining the counting process $\eta(t)$ which describes the indices associated with the progression of the diffusive trajectory through resetting events:
\begin{equation}
\eta(t) = 
\begin{cases}
    0 & t \leq t_1 \\
    m & t \in (t_m, t_{m+1}],~(1\le m\le n-1)\\
    n & t \in (t_n, T].
\end{cases}
\label{eq:eta_def}
\end{equation}
We note that statistically, $\eta(t)$ it is a stochastic process on the positive integers that increases by $1$ with Poissonian rate $r$ and satisfies the condition $\eta(T)=n$. To simplify this description, we write $\chi(t)=x_{\eta(t)}(t)$ or equivalently, $x_n(t)=\{\chi(t), \eta(t)\}$ as done in the main text, where $\chi(t)$ represents the purely diffusive trajectory that arrives at $x_n(t)$ at time $t$, $\eta(t)$ is a monotonically increasing (in forwards time) stochastic process on the integers $\{0, 1, \dots, n\}$ which describes how this trajectory is picked up by particle $x_1$ and passed down to $x_n$ by $n$ successive stochastic resetting events.

\begin{figure}
    \centering
    \includegraphics[width=\textwidth]{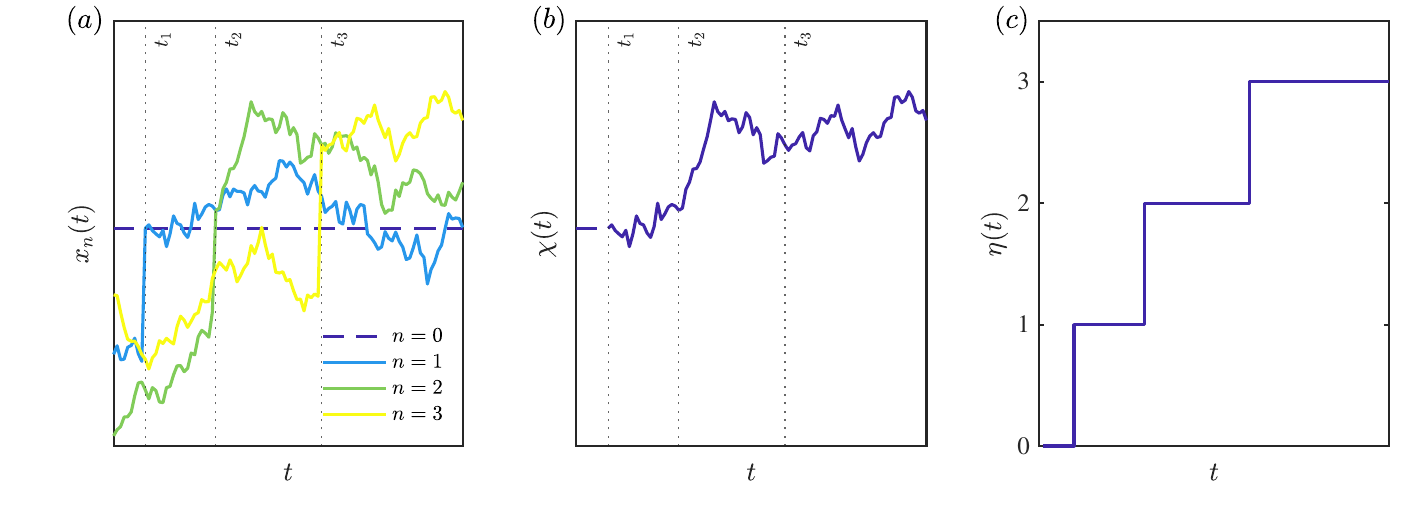}
    \caption{\textit{Example of mapping to diffusive trajectory} --- (a) Example realization of nested resetting process for $N=3$. (b) Building a diffusive trajectory $\chi(t)$ from the coupled microscopic dynamics as instructed from Eq.\,\eqref{eq:chi_def}. (c) The associated index from this trajectory, $\eta(t)$, such that $\chi(t) = x_{\eta(t)}(t)$ for all $t$.}
    \label{fig:difftraj}
\end{figure}

To derive the steady-state probability distribution for particle $n$, we first recall that the sum of $n$ independent and exponentially distributed random variables (drawn from the same distribution) is given by an Erlang-$n$ distribution. For resetting rate $r$, this takes the form
\begin{equation}
    E_n(t) = \frac{r^nt^{n-1}e^{-rt}}{(n-1)!}.
\end{equation}
This is exactly the distribution for the length of the time period $T-t_1$, namely the total time since the diffusive trajectory first left $0$. We now denote by $G_0(x, t)$ the diffusive Green's function in 1D, defined as
\begin{equation}
    G_0(x, t)=\frac{e^{-\frac{x^2}{4Dt}}}{\sqrt{4 \pi D t}}.
\end{equation}

To derive a distribution for the variable $\chi(T)$, we average over realizations of the timings $\{t_1,\dots, t_n\}$. In particular, $P_n^*(x)=\mathbb{P}(\chi(T)=x|\eta(T)=n)$ is a weighted average of the distributions $G_0(x, s)$, where the average is taken with respect to $s=T-t_1=\sum_{m=1}^n \tau_m$, where $s$ is Erlang-$n$ distributed and $\tau_m = t_{m+1} - t_{m}$. The steady-state distribution $P_n^*(x)$ can thus also be defined as
\begin{equation}
    \mathbb{P}(\chi(T)=x|\eta(T)=n)=\int_0^\infty ds \left[E_n(s) G_0(x, s)\right] = \frac{r^n}{\sqrt{4 \pi D} (n-1)!} \int_0^\infty ds\: s^{n-3/2}\exp\left[-rs-\frac{x^2}{4Ds} \right],
\end{equation}
which we re-write using the result \cite{Evans2020,Gradshtyn2014}
\begin{equation}
\int_0^{\infty} ds\: s^{n-3/2}\exp\left[-rs-\frac{x^2}{4Ds} \right] = 2 \left(\frac{x^2}{4Dr}\right)^{\frac{2n-1}{4}}K_{n-\frac{1}{2}}\left(\sqrt{\frac{r}{D}}x\right)
\end{equation}
and recalling that we defined $\alpha=\sqrt{r/D}$, we arrive at 
\begin{equation}
P_n^*(x) = \mathbb{P}(\chi(T)=x|\eta(T)=n) =\frac{\alpha}{\sqrt{\pi}(n-1)!}\biggl(\frac{\alpha|x|}{2}\biggr)^{n-1/2}K_{n-1/2}(\alpha |x|),
\end{equation}
which agrees exactly with Eq.\,(\textcolor{red}{6}) of the main text.

\section{Rederivation of positional moments}
\label{sec:positionalmoments}

This can also be determined from the mapping to a diffusive trajectory approach outlined in Sec.\,\ref{sec:difftraj}. Indeed, we write $x_n(T)=\{\chi(T), \eta(T)\}$ where, as above, $\chi(t)$ describes the diffusive trajectory that arrives at $x_n(T)$ at time $T$ (as defined in Eq.\,\eqref{eq:chi_def}) and $\eta(t)$ gives the indices associated with the diffusive trajectory at each point in time (as defined in Eq.\,\eqref{eq:eta_def}). 

Let $t_n$ again denote the last time before $T$ that $x_n$ reset to the position of $x_{n-1}$. This implies that $\chi(t_{n}) = x_{n-1}(t_{n})$, hence we can define $\delta \chi (T-t_n) = x_n(T) - x_{n-1}(t_{n})$ and find that
\begin{equation}
    x_n(T)^2 = (x_{n-1}(t_{n}) + \delta \chi(T-t_n))^2 = x_{n-1}(t_{n})^2 + 2 x_{n-1}(t_{n})\delta \chi(T-t_n) + \delta \chi(T-t_n)^2
\end{equation}
Noting that $\delta \chi(T-t_n)$ evolves independently of $x_{n-1}(t_{n})$, we average over realizations of the process and arrive at the following
\begin{equation}\label{eq:var_step}
    \langle x_n^2 \rangle = \langle x_{n-1}^2 \rangle + \langle \delta \chi^2 \rangle
\end{equation}
Finally, we evaluate $\langle \delta \chi^2 \rangle$ as the variance of a particle's displacement in $1$ resetting interval, namely over a single exponentially distributed waiting time, $T-t_n$. We recognize that this is exactly equivalent to $\langle x_1^2 \rangle$, thus $\langle x_n^2 \rangle = \langle x_{n-1}^2\rangle + \langle x_1^2 \rangle.$ As $n$ is here arbitrary, it follows that $\langle x_n^2\rangle = n\langle x_1^2\rangle$, agreeing with Eq.\,(9) in the main text. 

\section{Further details for the derivation of the steady-state correlations}
\label{sec:correlations}

\begin{figure}[b!]
    \centering
    \includegraphics[width=\textwidth]{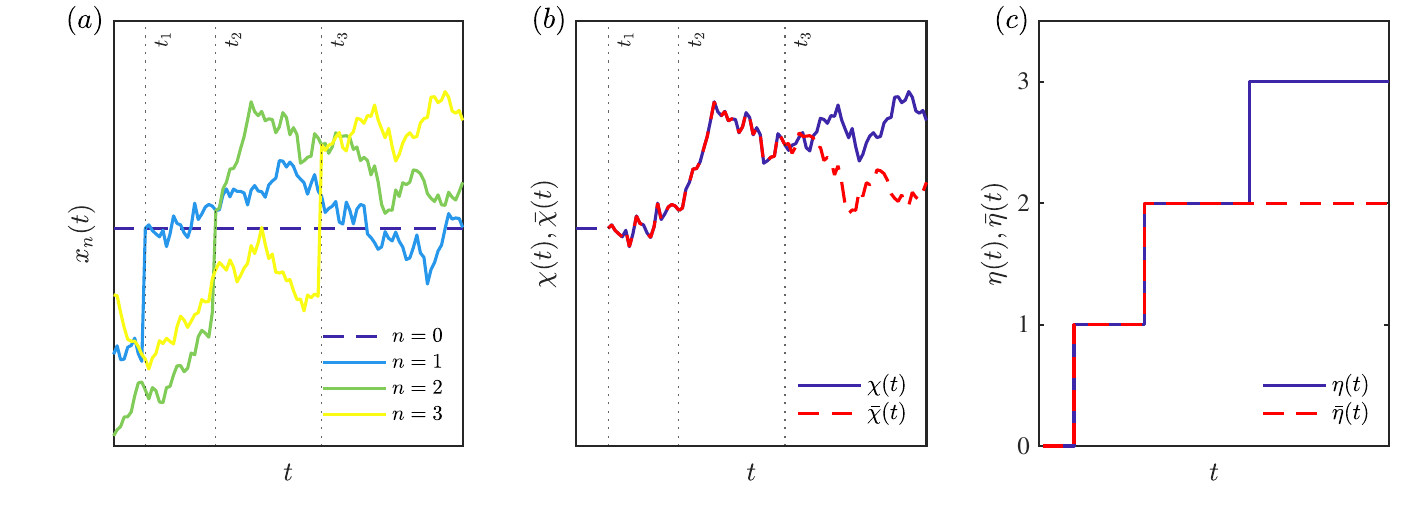}
    \caption{\textit{Diffusive trajectories for $x_2$ and $x_3$} --- As in Fig.\,\eqref{fig:difftraj}, we extract from the full process the fully diffusive trajectories from particle $n=2$ and $n=3$ (see original trajectories in (a)), this leads us to define $\chi(t)$ (resp. $\bar{\chi}(t)$) the diffusive trajectory corresponding to $x_3$ (resp. $x_2$). The associated counting processes $\eta(t)$ and $\bar{\eta}(t)$ are given in (c).}
    \label{fig:muldifftraj}
\end{figure}

In this section, we provide further details on the derivation of the correlations of the form $\langle x_{n}x_{\bar{n}}\rangle$, where $n$ and $\bar{n}>n$ are positive integers. To do so, we consider the two processes $x_n(t)=\{\chi(t), \eta(t)\}$ and $x_{\bar{n}}(t)=\{\bar{\chi}(t), \bar{\eta}(t)\}$, with $\bar{n} = n+j$, $j>0$. In the main text, we argued that the correlator between two particles on the resetting chain is entirely controlled by $t_\ell$ the last time before $T$ such that the $\eta(t_\ell) = \bar{\eta}(t_\ell)$ (see Fig.\,\ref{fig:muldifftraj}). 

 We then define, for $t>t_\ell$, $\delta \chi(t) = \chi(t)-\Upsilon$ and similarly $\delta \bar{\chi}(t)=\bar{\chi}(t)-\Upsilon$, where $\Upsilon \equiv \chi(t_\ell) = \bar{\chi}(t_\ell)$. We then write
\begin{equation}
    x_n(T) x_{\bar{n}}(T) = \chi (T) \bar{\chi}(T) = \Upsilon^2 + \left(\delta{\chi}(T)+\delta \bar{\chi}(T) \right)\Upsilon + \delta \chi(T) \delta \bar{\chi}(T).
\end{equation}
We now take the average over realizations of the process to arrive at the desired correlator. Importantly, $\delta X$ and $\delta \bar{X}$ are independent as there is no correlation between the diffusive motion of different particles in the system and $\eta(t) \ne \bar{\eta}(t)$ for $t>T_\ell$. Also, $\langle\delta \chi\rangle=0$ (and similarly for $\langle\delta \bar{\chi}\rangle$) as the diffusive process $\chi(t)$ is uncorrelated in time, which also implies that $\delta \chi$ and $\Upsilon$ are uncorrelated. The result is that only the first term survives the average and we derive the following simple form for the correlator
\begin{equation}
    \langle x_n x_{\bar{n}}\rangle = \langle \Upsilon^2\rangle.
\end{equation}

To evaluate the variance of $\Upsilon$, we consider the different values that $n_\ell$ could take, writing the variance as a sum weighted by the probability density function for $n_\ell$ (given $n$ and $\bar{n}$):
\begin{equation}
    \langle \Upsilon^2\rangle = \sum_{m=1}^{n}\langle \Upsilon^2\rangle_{n_\ell=m} \mathbb{P}(n_\ell=m~|~\{n, \bar{n}\}).
\end{equation}
(Note the sum stops at $n$ as $n<\bar{n}=n+j$.) The term $\langle \Upsilon^2\rangle_{n_\ell=m}$ is exactly the variance of the position of particle $x_{m}$: $\langle x_m^2\rangle$. We derived this above in Eq.\,(9) of the main text:
\begin{equation}
    \langle \Upsilon^2\rangle_{n_\ell=m} = \frac{2mD}{r}.
\end{equation}
The remainder of this section is devoted to evaluating the probability distribution for $\mathbb{P}(n_\ell=m~|~\{n, \bar{n}\})$. In what follows, we will write $\bar{n} = n+j$ with $j>0$. For brevity, we will denote the required probabilities by the coefficients
\begin{equation}
    \alpha_{m}^{n, n+j} \equiv \mathbb{P}(n_\ell=m |\{n, n+j\}).
\end{equation}
such that we can write the correlator in the form
\begin{equation}\label{eq:corr_expr}
    \langle x_n x_{n+j} \rangle = \sum_{m=1}^{n} \alpha_{m}^{n,n+j} \langle x_m^2 \rangle.
\end{equation}

\subsection{Mapping to a discrete-time random walk process}

We first show that the probabilities $\alpha_{m}^{n, n+j}$ can be determined entirely from $\eta(t)$ and $\bar{\eta}(t)$, namely they only consider the order of the resetting events and are independent of both the diffusive dynamics of the particles and the times between resetting events. To see this, we look again at large time $T$, where we have $\eta(T)=n$ and $\bar{\eta}(T)=n+j$. We then work backwards in time: we know that at some time $t_n<t$, particle $x_n$ will have reset to $x_{n-1}$ for the last time, so $\eta$ transitions from $n-1\rightarrow n$ at time $t_n$. Similarly, there exists a time $t_{n+j}$ at which $\bar{\eta}$ transitions from $n+j-1 \rightarrow n+j$. Given that the stochastic processes governing these resetting events are identical and independent, either $t_n<t_{n+j}$ or $t_n>t_{n+j}$ with equal probability. Let $t_{\rm max} = {\rm max}(t_n, t_{n+j})$, then just before $t_{\rm max}$ we know that either $(\eta, \bar{\eta})=(n-1, n+j)$ or $(\eta, \bar{\eta}) = (n, n+j-1)$ with equal probability. By iteration, we can calculate the probability that at some point in time prior to $T$, $(\eta(t), \bar{\eta}(t))=(n_1, n_2)$, given that $(\eta(T), \bar{\eta}(T))=(n, n+j).$ To be precise, $\alpha_m^{n, n+j}$ represents exactly the probability that $(\eta(t), \bar{\eta}(t))=(m, m)$ at some time $t$, but also crucially that there is no time $t'$ for which $(\eta(t'), \bar{\eta}(t'))=(m', m')$ for some $m'\in \{m+1, \dots, n-1, n\}$.

\begin{figure}
    \centering
    \includegraphics[width=\textwidth]{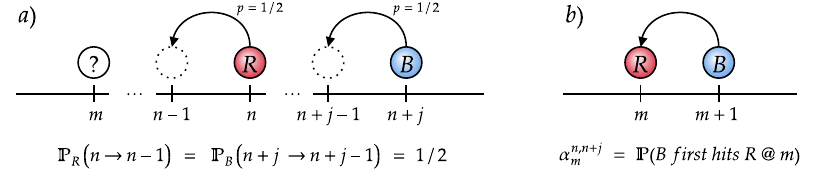}
    \caption{\textit{Schematic of DTRW process} --- In Section \ref{sec:correlations}, we calculate the steady-state correlations between particle positions of the form $\langle x_n x_{n+j}\rangle$ by mapping the problem to one of a first-hitting time of two random walkers: (a) Initially, one particle sits at site $n$ (here labelled $R$, colored red) and one at site $n+j$ ($B$, colored blue). To iterate the model in time, we choose one of the particles at random and move it to the left (from site $n'$ to $n'-1$). (b) The observable we calculate is exactly the probability that $B$ first reaches $R$ (that is, $B$ and $R$ are on the same site for the first time) at site number $m\leq n$.}
    \label{fig:schematic-DTRW}
\end{figure}

To calculate these probabilities, we map our problem to the following discrete-time random walk process on a 1D lattice with sites $s_i\in\mathbb{N}_{\geq0}$: Let one particle be placed initially at site $n$ and the other at site $n+j$. At each step in time, exactly one of the particles moves to the left (e.g.\,$n\rightarrow n-1$) and each particle moves with equal probability. Once a particle reaches $0$, it no longer moves [see Fig.\,\ref{fig:schematic-DTRW}]. Eventually, the particle initially at $n+j$ will hop on to the  site occupied by the particle initially at $n$: the site at which this occurs (for the first time) is exactly the site $m$ discussed above. Note that if the particle initially at $n$ reaches 0 before being caught by the other particle, then $m=0$. We now proceed to evaluate the probability that the particle initially at $n+j$ catches the particle initially at $n$ precisely at site $m$: this corresponds exactly with the quantity $\alpha_{m}^{n, n+j}$ detailed above.

\subsection{Calculating the probabilities $\alpha_{m}^{n, n+j}$}

In what follows, it will be useful to recall that the probability of $s$ successes in $t$ independent trials, each with a probability of success $p$ is given by a Binomial distribution which we denote by $\mathcal{B}(s, t, p)$ and is defined as
\begin{equation}\label{eq:binomial}
    \mathrm{B}(s, t, p) =  {t \choose s}\:p^s (1-p)^{t-s}.
\end{equation}

First we consider the case $m=n$. For the two random walkers to meet at site $n$, the particle initially at $n+j$ will need to do $j$ steps before the left particle takes any steps. Treating the movement of the right-most particle as the ``success" outcome (though the choice of left/right is irrelevant) with associated probability $p=1/2$ and defining the following shorthand notation 
\begin{equation}\label{eq:shorthand}
\mathcal{B}(s,t) = \mathrm{B}(s,t,1/2) = {t \choose s} \: 2^{-t},
\end{equation}
we write $\alpha_n^{n, n+j}$ in terms of $\mathcal{B}$ as
\begin{equation}
    \alpha_n^{n, n+j} = \mathcal{B}(j, j) =  2^{-j}.
\end{equation}

We now define the remaining coefficients in an iterative manner, proceeding first with $\alpha_{n-1}^{n, n+j}$. The probability that the two random walkers meet (not necessarily for the first time) at site $n-1$ is equal to the probability that in the first $j+2$ steps, the left particle only takes 1 step and the right particle takes $j+1$ steps. This is given by $\mathcal{B}(j+1, j+2)$ (where again successes correspond to movement of the right-most particle). Given that the two particles are together after $j+2$ steps, one of two possible mutually exclusive events has occurred:
\begin{enumerate}
    \item The two walkers have met \textit{for the first time} at site 
    $n-1$, which occurs with probability $\alpha_{n-1}^{n, n+j}$ (by definition);
    \item The two walkers met first at site $n$ (with probability $\alpha_{n}^{n, n+j}$) and then have each taken one step in the following two events (with probability $\mathcal{B}(1, 2)$).
\end{enumerate}

Equating the three probabilities, we derive a closed form equation for $\alpha_{n-1}^{n, n+j}$:
\begin{equation}
    \alpha_{n-1}^{n, n+j} = \mathcal{B}(j+1, j+2) - \alpha_n^{n, n+j} \mathcal{B}(1, 2)
\end{equation}

We can equate these path probabilities in the same manner for $\alpha_{n-k}^{n, n+j}$ where $k \in\{2,\dots, n-1\}$: this results in the sum
\begin{equation}\label{eq:alpha_coeff}
    \alpha_{n-k}^{n, n+j} = \mathcal{B}(k, j+2k) - \sum_{l=1}^{k} \alpha_{n-k+l}^{n, n+j} \mathcal{B}(l, 2l).
\end{equation}

Evaluating Eq.\,\eqref{eq:shorthand} for the probabilities $\mathcal{B}$, Eq.\,\eqref{eq:alpha_coeff} provides an exact, fully-analytic expression for all coefficients $\alpha_m^{n, n+j}$ and thus exact results for the correlations of the form $\langle x_nx_{n+j}\rangle$ through Eq.\,\eqref{eq:corr_expr} which constitutes the main result of this section. We show in Section \ref{app:jointpdf} that these correlations can also be calculated directly through the so-called last renewal approach widely used in the study of stochastic resetting processes. We find perfect agreement at small $n$ and $j$, beyond which the last renewal approach becomes computationally untractable. 

\begin{figure}[b!]
    \centering
    \includegraphics[width=\textwidth]{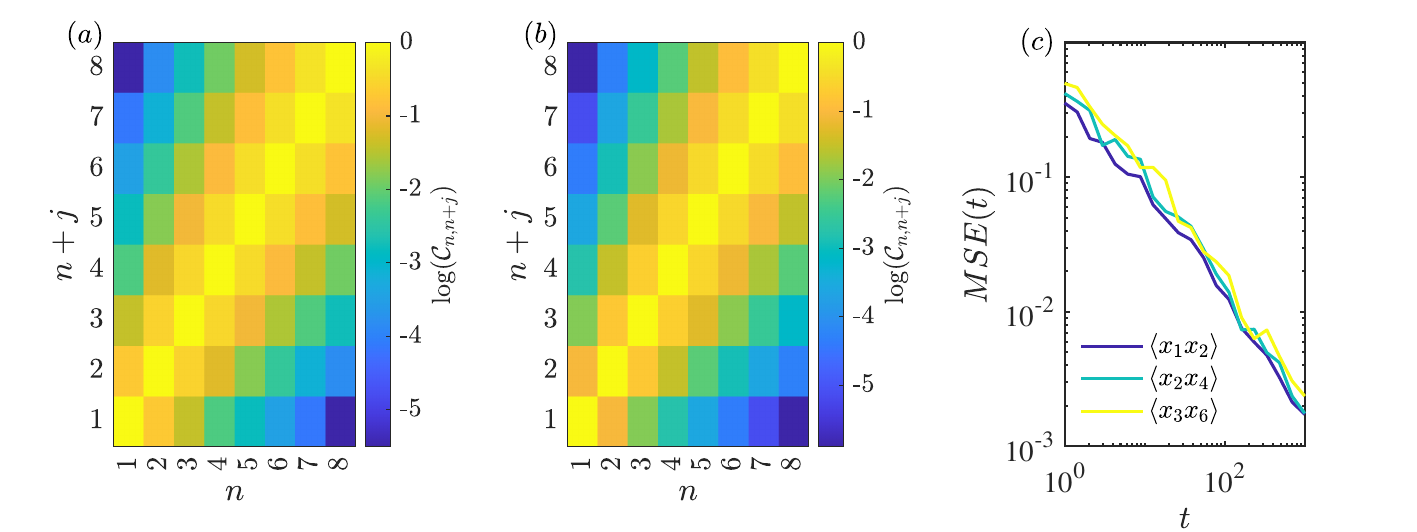}
    \caption{\textit{Exact correlations in nested resetting processes} --- We compare results for the correlations of the particle positions of the form $\langle x_nx_{n+j}\rangle$ from (a) numerical simulations and (b) analytic results derived in Section \ref{sec:correlations}. (c) We plot the mean-squared error, $MSE(t)$ as defined in Eq.\,\eqref{eq:mse}, demonstrating that $MSE \sim t^{\beta}$ where $\color{red}\beta<0$ and $t$ is the simulation length, ensuring convergence to our analytic result for sufficiently long simulations, thus the exactness of our result.}
    \label{fig:heatmaps}
\end{figure}

\subsection{Numerical verification for small $n$ and $j$}

In Figure \ref{fig:heatmaps}, we demonstrate that our results agree with numerical calculations of the correlators from microscopic realizations of the process, confirming the exactness of our analytic result. We consider the re-scaled correlation function $\mathcal{C}_{n, n+j}$ defined as 
\begin{equation}\label{eq:resclcorr}
    \mathcal{C}_{n, n+j} = \frac{\langle x_{n}x_{n+j}\rangle}{\sqrt{\langle x_n^2\rangle \langle x_{n+j}^2\rangle}}.
\end{equation}
In particular, we show that the mean-squared error defined as
\begin{equation}\label{eq:mse}
    \mathrm{MSE}(t) = |\mathcal{C}_{n, n+j} - \mathcal{C}^{\rm num}_{n, n+j}(t)  |^2,
\end{equation}
where $\mathcal{C}^{\rm num}_{n, n+j}(t)$ is the numerically measured correlation function from a simulation of length $t$, decays monotonically as a function of time. More specifically, we show numerically in Fig.\,\ref{fig:heatmaps}(c) that $\mathrm{MSE} \sim t^{\beta}$ where $\beta < 0$, and thus that the mean-squared error tends to zero as the simulation length increases indefinitely. 

\section{Steady-state correlators from a last-renewal approach}
\label{app:jointpdf}

In the following section, we re-derive the two-particle correlators using the joint $n$-particle distribution. The latter can naturally be formulated using a last renewal approach as
\begin{multline}
    \mathcal{P}_{n}(\{x_i\},t|\{x_{i}(0)\}) = e^{-rt}G_0(x_n, t|x_{n}(0))\mathcal{P}_{n}(\{x_i\}\},t|\{x_{i}(0)\}) + \int_{-\infty}^{+\infty} d\tilde{x}_1\dots d\tilde{x}_{n-1}\int^{t}_{0} d\tau re^{-r\tau} G_0(x_n, \tau|\tilde{x}_{n-1})\\
    \times \mathcal{P}_{n-1}(\{\tilde{x}_i\},t-\tau| \{x_{i}(0)\}) \mathcal{P}_{n-1}(\{x_i\},\tau|\{\tilde{x}_i\}),
\end{multline}
which at steady-state ($t \to \infty$) leads to 
\begin{equation}
    \mathcal{P}^{*}_{n}(\{x_i\}) = \int_{-\infty}^{+\infty} d\tilde{x}_1\dots d\tilde{x}_{n-1}\int^{+\infty}_{0} d\tau re^{-r\tau} G_0(x_n, \tau|{\tilde{x}}_{n-1})
    \mathcal{P}^{*}_{n-1}(\{\tilde{x}_i\}) \mathcal{P}_{n-1}(\{x_i\},\tau|\{\tilde{x}_i\}),
\end{equation}

This can be marginalized to produce the joint distribution for any pair of particles, something that can subsequently be used to calculate two-particle correlators.

\subsection{Calculation of $\langle x_2 x_1\rangle$}

We now focus on the calculation of the most trivial two particle correlator, $\langle x_2 x_1 \rangle$. Using the joint-pdf as formulated above one can calculate the correlator as follows
\begin{flalign}
    \langle x_2 x_1 \rangle &= \int^{+\infty}_{-\infty} dx_1 dx_2 \, x_2 x_1 \mathcal{P}_{2}^{*}(x_1,x_2)\nonumber\\
    &= \int^{+\infty}_{-\infty} d\tilde{x} \int^{+\infty}_{0} d\tau \, re^{-2r\tau} \tilde{x}^2 P^{*}_{1}(\tilde{x})\nonumber\\
    &= \frac{1}{2} \langle x_1^2 \rangle,
\end{flalign}
which agrees with our above algorithm for calculating correlators, where we find that
\begin{align}
    \langle x_2 x_1 \rangle = \alpha_{1}^{1,2} \langle x_{1}^2 \rangle,
\end{align}
with
\begin{align}
    \alpha_1^{1, 2} = \mathcal{B}(1, 1) =  \frac{1}{2}.
\end{align}

\subsection{Calculation of $\langle x_3 x_2\rangle$ and $\langle x_3 x_1\rangle$}

In this subsection, we calculate $\langle x_3 x_2\rangle$ and $\langle x_3 x_1\rangle$, both of which can be calculated from the 3 particle joint-pdf. We find
\begin{flalign}
    \langle x_3 x_2 \rangle &= \int^{+\infty}_{-\infty} dx_1 dx_2 dx_3 \, x_3 x_2 \mathcal{P}_{3}^{*}(x_1,x_2,x_3)\nonumber\\
    &= \int^{+\infty}_{-\infty} d\tilde{x}_1 d\tilde{x}_2 \int^{+\infty}_{0} d\tau \, [re^{-2r\tau} \tilde{x}_{2}^2 + r^{2} \tau e^{-2r\tau} \tilde{x}_1 \tilde{x}_2 ]\mathcal{P}^{*}_{2}(\tilde{x}_1,\tilde{x}_2)\nonumber\\
    &= \frac{1}{2}\langle x_{2}^{2}\rangle + \frac{1}{8}\langle x_{1}^{2}\rangle,\\
    \langle x_3 x_1 \rangle &= \int^{+\infty}_{-\infty} dx_1 dx_2 dx_3 \, x_3 x_1 \mathcal{P}_{3}^{*}(x_1,x_2,x_3)\nonumber\\
    &= \int^{+\infty}_{-\infty} d\tilde{x}_1 d\tilde{x}_2 \int^{+\infty}_{0} d\tau \, [re^{-r\tau} - re^{-2r\tau}]\tilde{x}_1 \tilde{x}_2 \mathcal{P}^{*}_{2}(\tilde{x}_1,\tilde{x}_2)\nonumber\\
    &= \frac{1}{4}\langle x_{1}^2 \rangle.
\end{flalign}
By our algorithm in Section 5, we find that
\begin{align}
    \langle x_3 x_2 \rangle &= \alpha_{2}^{2,3}\langle x_{2}^{2} \rangle + \alpha_{1}^{2,3}\langle x_{1}^{2}\rangle,\\
    \langle x_3 x_1 \rangle &= \alpha_{1}^{1,3}\langle x_{1}^{2} \rangle
\end{align}
with
\begin{align}
    \alpha_{2}^{2,3} &= \mathcal{B}(1, 1) =  \frac{1}{2},\\
    \alpha_{1}^{2,3} &= \mathcal{B}(2, 3) - \frac{1}{2}\mathcal{B}(1, 2, 1/2) = \frac{1}{8},\\
    \alpha_{1}^{1,3} &= \mathcal{B}(2, 2) = \frac{1}{4},
\end{align}
showing agreement between the algorithm and calculation for values of $n$ and $j$ for which the last renewal method is tractable straightforwardly. 
	

%